\documentclass[aps,prl,reprint,showpacs,superscriptaddress]{revtex4-1}

\usepackage{physics}
\usepackage[dvipdfmx]{graphicx}
\usepackage{amsmath,amssymb}
\usepackage{xparse}
\usepackage{color}
\usepackage{comment}
\usepackage{color}
\usepackage[normalem]{ulem}
\usepackage{cancel}

\def\HFI{H_{\mathrm{FI}}}
\def\HM{H_{\mathrm{M}}}
\def\Hex{H_{\mathrm{ex}}}
\def\HZ{H_{\mathrm{Z}}}
\def\HT{H_{\mathrm{T}}}
\def\la{\langle}
\def\ra{\rangle}
\def\vr{\vb*{r}}
\def\vk{\vb*{k}}
\def\vq{\vb*{q}}
\def\kbt{k_{\mathrm{B}}T}

\def\HBdG{H_{\mathrm{BdG}}}

\def\vp{\vb*{\psi}}
\def\hPBdG{\vb*{\Psi}_{\mathrm{BdG}}}
\def\EF{E_{\mathrm{F}}}
\def\kF{k_{\mathrm{F}}}

\def\kbt{k_{\mathrm{B}}T}

\def\ehk{\varepsilon_{\hat{\vb*{k}}}}
\def\kp{\vb*{k}_{\parallel}}
\def\qp{\vb*{q}_{\parallel}}
\def\rp{\vb*{r}_{\parallel}}
\def\hk{\hat{k}}
\def\hP{\Psi}
\def\hp{\psi}

\def\uvec{\mqty(
        u_{\to} \\
        u_{\gets} \\
        v_{\to} \\
        v_{\gets}
        )}
\def\pkp{\phi_{\vb*{k}_{\parallel}}}

\begin{document}

\title{Dynamical Majorana Ising spin response in a topological superconductor-magnet hybrid by microwave irradiation}

\author{Yuya Ominato}
\affiliation{%
Kavli Institute for Theoretical Sciences, University of Chinese Academy of Sciences, Beijing, 100190, China.
}%
\affiliation{%
Waseda Institute for Advanced Study, Waseda University, Shinjuku, Tokyo 169-8050, Japan.
}%
\author{Ai Yamakage}
\affiliation{Department of Physics, Nagoya University, Nagoya 464-8602, Japan}
\author{Mamoru Matsuo }
\affiliation{%
Kavli Institute for Theoretical Sciences, University of Chinese Academy of Sciences, Beijing, 100190, China.
}%
\affiliation{%
CAS Center for Excellence in Topological Quantum Computation, University of Chinese Academy of Sciences, Beijing 100190, China
}%
\affiliation{%
Advanced Science Research Center, Japan Atomic Energy Agency, Tokai, 319-1195, Japan
}%
\affiliation{%
RIKEN Center for Emergent Matter Science (CEMS), Wako, Saitama 351-0198, Japan
}%

\date{\today}

\begin{abstract}
We study a dynamical spin response of surface Majorana modes in a topological superconductor-magnet hybrid under microwave irradiation. We find a method to toggle between dissipative and non-dissipative Majorana Ising spin dynamics by adjusting the external magnetic field angle and the microwave frequency. This reflects the topological nature of the Majorana modes, enhancing the Gilbert damping of the magnet, thereby, providing a detection method for the Majorana Ising spins. Our findings illuminate a magnetic probe for Majorana modes, paving the path to innovative spin devices.
\end{abstract}

\maketitle 

{\it Introduction.---}The quest for Majoranas within matter stands as one of the principal challenges in the study of condensed matter physics, more so in the field of quantum many-body systems \cite{Wilczek2009}.
The self-conjugate nature of Majoranas leads to peculiar electrical characteristics that have been the subject of intensive research, both theoretical and experimental \cite{Yazdani2023}.
In contrast, the focus of this paper lies on the magnetic properties of Majoranas, specifically the Majorana Ising spin \cite{Chung2009, Sato2009-kv, Nagato2009-ht, Shindou2010-jl,Shiozaki2014-dh, Xiong2017}. A distinctive characteristic of Majorana modes, appearing as a surface state in topological superconductors (TSC), is its exceedingly strong anisotropy, which makes it behave as an Ising spin. In particular, this paper proposes a method to explore the dynamical response of the Majorana Ising spin through the exchange interaction at the magnetic interface, achieved by coupling the TSC to a ferromagnet with ferromagnetic resonance (FMR) (as shown in Fig.1 (a)).

FMR modulation in a magnetic hybrid system has attracted much attention as a method to analyze spin excitations in thin-film materials attached to magnetic materials \cite{Qiu2016,Han2020spin}. Irradiating a magnetic material with microwaves induces dynamics of localized spin in magnetic materials, which can excite spins in adjacent thin-film materials via the magnetic proximity effect. This setup is called spin pumping, and has been studied intensively in the field of spintronics as a method of injecting spins through interfaces \cite{tserkovnyak2002enhanced,tserkovnyak2005nonlocal}. Recent studies have theoretically proposed that spin excitation can be characterized by FMR in hybrid systems of superconducting thin films and magnetic materials \cite{inoue2017,Kato2019,silaev2020a,silaev2020b,Ominato2022a,Ominato2022b}. Therefore, it is expected to be possible to analyze the dynamics of surface Majorana Ising spins using FMR in hybrid systems.

\begin{figure}[h]
\begin{center}
\includegraphics[width=1\hsize]{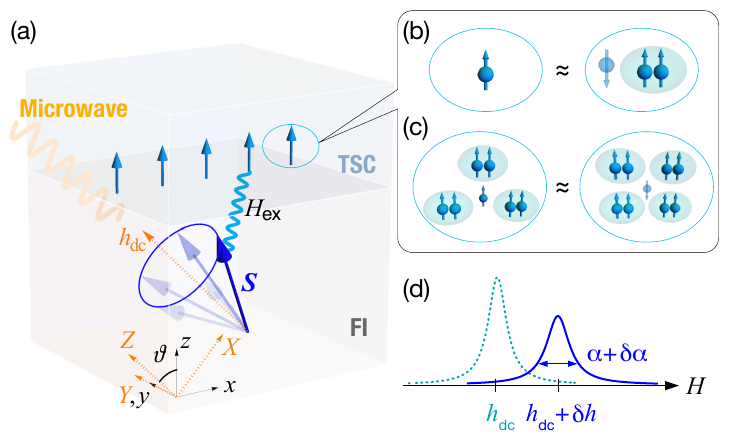}
\end{center}
\caption{
(a) The TSC-FI hybrid schematic reveals how, under resonance frequency microwave irradiation, localized spins commence precessional motion, consequently initiating the dynamical Majorana Ising spin response at the TSC interface. (b) In the TSC context, the liaison between a spin-up electron and a spin-down hole with the surrounding sea of spin-triplet Cooper pairs drastically modulate their properties; notably, a spin-down hole can engage with a spin-triplet Cooper pair, thereby inheriting a negative charge. (c) Notably, spin-triplet Cooper pairs amass around holes and scatter around electrons, thereby eroding the rigid distinction between the two. (d) The interplay between the Majorana mode and the localized spin manipulates the FMR spectrum, triggering a frequency shift and linewidth broadening.
}
\label{fig_system}
\end{figure}

In this work, we consider a TSC-ferromagnetic insulator (FI) hybrid system as shown in Fig.\ \ref{fig_system} (a). The FMR is induced by microwave irradiation on the FI. At the interface between the TSC and the FI, the surface Majorana modes interact with the localized spins in the FI.
As a result, the localized spin dynamics leads to the dynamical Majorana Ising spin response (DMISR), which means the Majorana Ising spin density is dynamically induced, and it is possible to toggle between dissipative and non-dissipative Majorana Ising spin dynamics by adjusting the external magnetic field angle and the microwave frequency.
Furthermore, the modulation of the localized spin dynamics due to the interface interaction leads to a frequency shift and a linewidth broadening, which reflect the properties of the Majorana Ising spin dynamics.
This work proposes a setup for detecting Majorana modes and paves the way for the development of quantum computing and spin devices using Majoranas.

{\it Model.---}We introduce a model Hamiltonian $H$ consisting of three terms
\begin{align}
    H=\HM+\HFI+\Hex.
\end{align}
The first, second, and third terms respectively describe the surface Majorana modes on the TSC surface, the bulk FI, and the proximity-induced exchange coupling. Our focus is on energy regions significantly smaller than the bulk superconducting gap. This focus allows the spin excitation in the TSC to be well described using the surface Majorana modes. The subsequent paragraphs provide detailed explanations of each of these three terms.

The first terms $\HM$ describes the surface Majorana modes,
\begin{align}
    \HM=\frac{1}{2}\int d\vr
    \vp^{\mathrm{T}}(\vr)
    \qty(\hbar v\hat{k}_y\sigma^x-\hbar v\hat{k}_x\sigma^y)
    \vp(\vr),
    \label{eq_Majorana}
\end{align}
where $\vr=(x,y)$, $\hat{\vk}=(-i\partial_x,-i\partial_y)$, $v$ is a constant velocity, and $\vb*{\sigma}=(\sigma^x,\sigma^y,\sigma^z)$ are the Pauli matrices.
The two component Majorana field operator is given by $\vp(\vr)=\qty(\psi_{\to}(\vr),\psi_{\gets}(\vr))^{\mathrm{T}}$, with the spin quantization axis along the $x$ axis.
The Majorana field operators satisfy the Majorana condition
$\psi_{\sigma}(\vr)=\psi_{\sigma}^\dagger(\vr)$ and the anticommutation relation
$\{\psi_{\sigma}(\vr),\psi_{\sigma^\prime}(\vr)\}=\delta_{\sigma\sigma^\prime}\delta(\vr-\vr^\prime)$ where $\sigma,\sigma^\prime=\to,\gets$.
We can derive $\HM$ by using surface-localized solutions of the BdG equation based on the bulk TSC Hamiltonian. The details of the derivation of $\HM$ are provided in the Supplemental Material \cite{sm}.

A notable feature of the surface Majorana modes is that the spin density is Ising like, which we call the Majorana Ising spin \cite{Chung2009, Sato2009-kv, Nagato2009-ht, Shindou2010-jl,Shiozaki2014-dh, Xiong2017}.
The feature follows naturally from the Majorana condition and the anticommutation relation.
The Majorana Ising spin density operator is given by $\vb*{s}(\vr):=\vp^{\mathrm{T}}(\vr)\qty(\vb*{\sigma}/2)\vp(\vr)=(0,0,-i\psi_{\to}(\vr)\psi_{\gets}(\vr))$ (See the Supplemental Material for details \cite{sm}).
The anisotropy of the Majorana Ising spin is the hallmark of the surface Majorana modes on the TSC surface.

The second term $\HFI$ descries the bulk FI and is given by the ferromagnetic Heisenberg model,
\begin{align}
    \HFI=
    &-\mathcal{J}\sum_{\la n,m\ra}\vb*{S}_n\cdot\vb*{S}_m
    -\hbar\gamma h_{\mathrm{dc}}\sum_n S_n^Z,
\end{align}
where $\mathcal{J}>0$ is the exchange coupling constant, $\vb*{S}_n$ is the localized spin at site $n$, $\la n,m\ra$ means summation for nearest neighbors, $\gamma$ is the electron gyromagnetic ratio, and $h_{\mathrm{dc}}$ is the static external magnetic field.
We consider the spin dynamics of the localized spin under microwave irradiation, applying the spin-wave approximation. This allows the spin excitation to be described by a free bosonic operator, known as a magnon \cite{Holstein1940}.

The third term $\Hex$ represents the proximity exchange coupling at the interface between the TSC and the FI,
\begin{align}
    &\Hex=-\int d\vr\sum_n
    J(\vr,\vr_n)\vb*{s}(\vr)\cdot\vb*{S}_n=\HZ+\HT, \\
    &\HZ=-\cos\vartheta\int d\vr\sum_n
    J(\vr,\vr_n)s^z(\vr)S^Z_n, \\
    &\HT=-\sin\vartheta\int d\vr\sum_n
    J(\vr,\vr_n)
    s^z(\vr)S_n^X,
\end{align}
where the angle $\vartheta$ is shown in Fig.\ \ref{fig_system} (a).
$H_{\mathrm{Z}}$ is the coupling along the precession axis and $H_{\mathrm{T}}$ is the coupling perpendicular to the precession axis.
In our setup, $\HZ$ leads to gap opening of the energy spectrum of the surface Majorana modes and $\HT$ gives the DMISR under the microwave irradiation.

{\it Dynamical Majorana Ising spin response.---}We consider the microwave irradiation on the FI. The coupling between the localized spins and the microwave is given by
\begin{align}
    V(t)=-\hbar\gamma h_{\mathrm{ac}}
    \sum_n\qty(S^X_n\cos{\omega t}-S^Y_n\sin{\omega t}),
\end{align}
where $h_{\mathrm{ac}}$ is the microwave amplitude, and $\omega$ is the microwave frequency.
The microwave irradiation leads to the precessional motion of the localized spin.
When the frequency of the precessional motion and the microwave coincide, the FMR occurs.
The FMR leads to the DMISR via the exchange interaction.
The DMISR is characterized by the dynamic spin susceptibility of the Majorana modes, $\tilde{\chi}^{zz}(\vq,\omega)$, defined as
\begin{align}
    \tilde{\chi}^{zz}(\vq,\omega)
    :=\int d{\vr}e^{-i\vq\cdot\vr}\int dt e^{i(\omega+i0)t}
    \chi^{zz}(\vr,t),
\end{align}
where $\chi^{zz}(\vr,t):=-(L^2/i\hbar)\theta(t)\la\qty[s^z(\vr,t),s^z(\vb*{0},0)]\ra$ with the interface area $L^2$ and the spin density operator in the interaction picture, $s^z(\vr,t)=e^{i(\HM+\HZ)t/\hbar}s^z(\vr)e^{-i(\HM+\HZ)t/\hbar}$.
For the exchange coupling, we consider configuration average and assume
$\la\sum_nJ(\vr,\vr_{n})\ra_{\mathrm{ave}}=J_1$
, which means that $\HZ$ is treated as a uniform Zeeman like interaction and the interface is specular \footnote{In a realistic sample, correction terms would be added due to interface roughness \cite{Ominato2022a,Ominato2022b}. Even with such correction terms, the characteristic angular dependence is expected to be preserved, since the Majorana Ising spin property of having spin density only in the perpendicular direction is maintained.}.
Using eigenstates of Eq.~(\ref{eq_Majorana}) and after a straightforward calculation, the uniform spin susceptibility is given by
\begin{align}
    &\tilde{\chi}^{zz}(\vb{0},\omega) \notag \\
    &=-\sum_{\vk,\lambda}
    \abs{\langle\vk,\lambda|\sigma^z|\vk,-\lambda\rangle}^2
    \frac{f(E_{\vk,\lambda})-f(E_{\vk,-\lambda})}{2E_{\vk,\lambda}+\hbar\omega+i0}, \notag \\
    &\to-\int dED(E)\frac{E^2-M^2}{2E^2}
    \frac{f(E)-f(-E)}{2E+\hbar\omega+i0},
\end{align}
where $|\vk,\lambda\rangle$ is an eigenstate of $\HM$ with eigenenergy $E_{\vk,\lambda}=\lambda\sqrt{(\hbar vk)^2+M^2}$, ($\lambda=\pm$). $M=J_1S\cos\vartheta$ is the Majorana gap, $f(E)=1/(e^{E/\kbt}+1)$ is the Fermi distribution function, and $D(E)$ is the density of states given by
\begin{align}
    D(E)=\frac{L^2}{2\pi(\hbar v)^2}\abs{E}\theta\qty(\abs{E}-\abs{M}),
\end{align}
with the Heaviside step function $\theta(x)$.
It is important to note that the behavior of the uniform spin susceptibility is determined by the interband contribution, which is proportional to the Fermi distribution function, i.e., the contribution of the occupied states. This mechanism is similar to the Van Vleck paramagnetism \cite{ashcroft1976}. The contribution of the occupied states often plays a crucial role in topological responses \cite{murakami2006}.

\begin{figure}[t]
\begin{center}
\includegraphics[width=1\hsize]{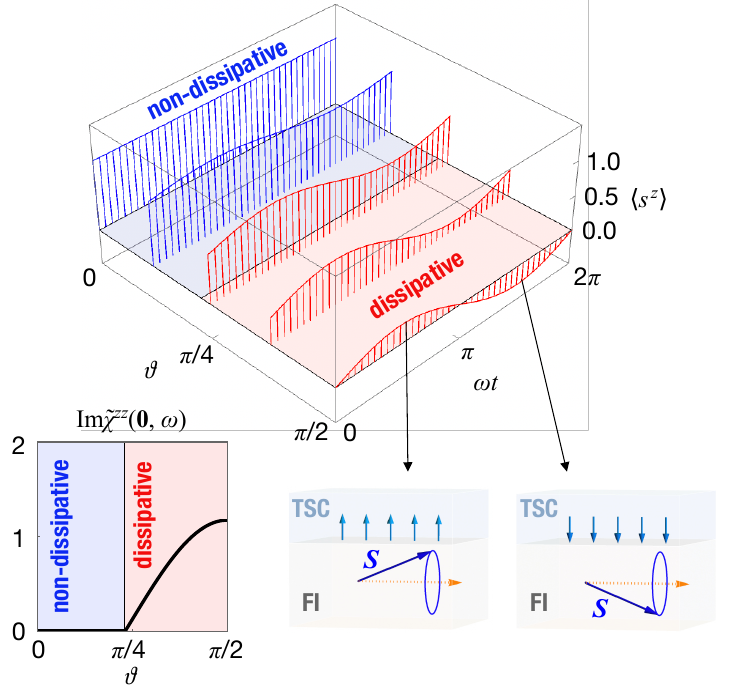}
\end{center}
\caption{The induced Ising spin density, with a unit $\tilde{\chi}^{zz}_0(\vb*{0},0)J_1S$, is presented as a function of $\omega t$ and $\vartheta$.
The frequency and temperature are set to $\hbar\omega/J_1S=1.5$ and $\kbt/J_1S=0.1$, respectively.
The coefficient, $h_{\mathrm{ac}}/\alpha h_{\mathrm{dc}}$, is set to $0.3$.
The static Majorana Ising spin density arises from $\HZ$. When the precession axis deviates from the direction perpendicular to the interface, the precessional motion of the localized spins results in the dynamical Majorana Ising spin response (DMISR). Energy dissipation due to the DMISR is zero for small angles $\vartheta$ as the Majorana gap exceeds the magnon energy. However, once the magnon energy overcomes the Majorana gap, the energy dissipation becomes finite.
Therefore, one can toggle between dissipative and non-dissipative DMISR by adjusting $\vartheta$.
}
\label{fig_DMISR}
\end{figure}

Replacing the localized spin operators with their statistical average values, we find the induced Majorana Ising spin density, to the first order of $J_1S$, is given by
\begin{align}
    \int d\vr\expval{s^z(\vr,t)}
    &=\tilde{\chi}^{zz}_0(\vb*{0},0)J_1S\cos\vartheta \notag \\
    &\hspace{-10mm}
    +\mathrm{Re}\qty[\tilde{\chi}^{zz}_0(\vb*{0},\omega)]\frac{ h_{\mathrm{ac}}}{\alpha h_{\mathrm{dc}}}J_1S\sin\vartheta\sin\omega t,
    \label{eq_Ising_spin}
\end{align}
where $\tilde{\chi}^{zz}_0(\vb*{0},0)$ is the spin susceptibility for $M=0$.
The first term originates from $\HZ$ and gives a static spin density, while the second term originates from $\HT$ and gives a dynamic spin density.
Figure \ref{fig_DMISR} shows the induced Ising spin density as a function of time at several angles.
As shown in Eq.\ (\ref{eq_Ising_spin}), the Ising spin density consists of the static and dynamic components.
The dynamic component is induced by the precessional motion of the localized spin, which means one can induce the DMISR using the dynamics of the localized spin.

The inset in Fig.\ \ref{fig_DMISR} shows $\mathrm{Im}\tilde{\chi}^{zz}(\vb*{0},\omega)$ as a function of $\vartheta$ at a fixed frequency.
When the frequency $\hbar \omega$ is smaller than the Majorana gap, $\mathrm{Im}\tilde{\chi}^{zz}(\vb*{0},\omega)$ is zero.
Once the frequency overcomes the Majorana gap, $\mathrm{Im}\tilde{\chi}^{zz}(\vb*{0},\omega)$ becomes finite.
The implications of these behaviors are that if the magnon energy is smaller than the Majorana gap, there is no energy dissipation due to the DMISR. However, once the magnon energy exceeds the Majorana gap, finite energy dissipation associated with the DMISR occurs at the surface of the TSC.
Therefore, one can toggle between dissipative and non-dissipative Majorana Ising spin dynamics by adjusting the precession axis angle and the microwave frequency.

{\it FMR modulation.---}The retarded component of the magnon Green's function is given by
$G^R(\vr_n,t)=-(i/\hbar)\theta(t)\la[S^+_{n}(t),S^-_{0}(0)]\ra$ with the interaction picture $S^{\pm}_n(t)=e^{i\HFI t/\hbar}S^{\pm}_ne^{-i\HFI t/\hbar}$.
The FMR signal is characterized by the spectral function defined as
\begin{align}
    A(\vq,\omega):=-\frac{1}{\pi}
    \mathrm{Im}
    \qty[
        \sum_n
        e^{-i\vq\cdot\vr_n}
        \int dt e^{i(\omega+i0)t}
        G^R(\vr_n,t)
    ].
\end{align}
For uniform external force, the spectral function is given by
\begin{align}
    A(\vb{0},\omega)=
    \frac{2S}{\hbar}
    \frac{1}{\pi}
    \frac{(\alpha+\delta\alpha)\omega}{\qty[\omega-\gamma (h_{\mathrm{dc}}+\delta h)]^2+\qty[(\alpha+\delta\alpha)\omega]^2}.
\end{align}
The peak position and width of the FMR signal is given by $h_{\mathrm{dc}}+\delta h$ and $\alpha+\delta\alpha$, respectively.
$h_{\mathrm{dc}}$ and $\alpha$ correspond to the peak position and the linewidth of the FMR signal of the FI alone.
$\delta h$ and $\delta\alpha$ are the FMR modulations due to the exchange interaction $\HT$.
We treat $\HM+\HFI+\HZ$ as an unperturbed Hamiltonian and $\HT$ as a perturbation.
In this work, we assume the specular interface, where the coupling $J(\vr,\vr_n)$ is approximated as
$\Big\la\sum_{n,n^\prime}{J(\vr,\vr_{n})J(\vr^{\prime},\vr_{n^\prime})}\Big\ra_{\mathrm{ave}}=J_1^2$.
The dynamics of the localized spins in the FI is modulated due to the interaction between the localized spins and the Majorana Ising spins.
In our setup, the peak position and the linewidth of the FMR signal are modulated and the FMR modulation is given by
\begin{align}
    &\delta h
    =\sin^2\vartheta
    \frac{SJ^2_1}{2N\gamma\hbar}
    \mathrm{Re}\tilde{\chi}^{zz}(\vb*{0},\omega), \label{eq_dh} \\
    &\delta\alpha=\sin^2\vartheta\frac{SJ^2_1}{2N\hbar\omega}
    \mathrm{Im}\tilde{\chi}^{zz}(\vb*{0},\omega),
    \label{eq_da}
\end{align}
where $N$ is the total number of sites in the FI.
These formulas were derived in the study of the FMR in magnetic multilayer systems including superconductors.
One can extract the spin property of the Majorana mode from the data on $\delta h$ and $\delta\alpha$.
Because of the Ising spin anisotropy, the FMR modulation exhibits strong anisotropy, where the FMR modulation is proportional to $\sin^2\vartheta$.

\begin{figure}
\begin{center}
\includegraphics[width=1\hsize]{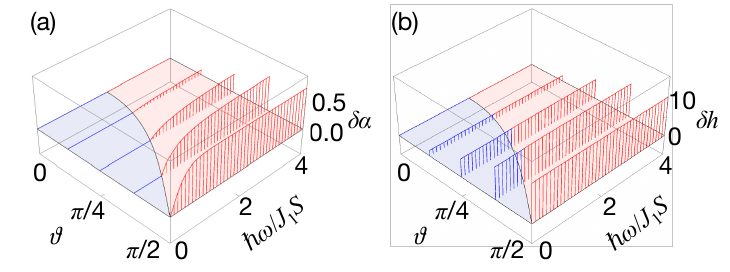}
\end{center}
\caption{
The temperature is set to $\kbt/J_1S=0.1$. (a) The damping modulation $\delta\alpha$ only becomes finite when the magnon energy exceeds the Majorana gap; otherwise, it vanishes. This behavior corresponds to the energy dissipation of the Majorana Ising spin. (b) The peak shift is finite, except for $\vartheta=0$, and is almost independent of $\omega$. This behavior resembles the spin response observed in strongly spin-orbit coupled band insulators, where the interband contribution to spin susceptibility results in a finite spin response, even within the energy gap.}
\label{fig_FMR_modulation}
\end{figure}

Figure \ref{fig_FMR_modulation} shows the FMR modulations (a) $\delta\alpha$ and (b) $\delta h$. The FMR modulation at a fixed frequency increases with angle $\vartheta$ and reaches a maximum at $\pi/2$, as can be read from Eqs.\ (\ref{eq_dh}) and (\ref{eq_da}).
When the angle $\vartheta$ is fixed and the frequency $\omega$ is increased, $\delta\alpha$ becomes finite above a certain frequency at which the energy of the magnon coincides with the Majorana gap.
When $\vartheta<\pi/2$ and $\hbar\omega\approx 2M$,
$\delta\alpha$ linearly increases as a function of $\omega$ just above the Majorana gap.
The localized spin damping is enhanced when the magnon energy exceeds the Majorana gap.
At $\vartheta=\pi/2$ and $\omega\approx0$, the Majorana gap vanishes and $\delta\alpha$ is proportional to $\omega/T$.
In the high frequency region $\hbar\omega/J_1S\gg1$, $\delta\alpha$ converges to its upper threshold.

The frequency shift $\delta h$ is almost independent of $\omega$ and has a finite value even in the Majorana gap. This behavior is analogous to the interband contribution to the spin susceptibility in strongly spin-orbit coupled band insulators, and is due to the fact that the effective Hamiltonian of the Majorana modes includes spin operators. It is important to emphasize that although the Majorana modes have spin degrees of freedom, only the $z$ component of the spin density operator is well defined. This is a hallmark of Majorana modes, which differs significantly from electrons in ordinary solids.
Note that $\delta h$ is proportional to the energy cutoff, which is introduced to converge energy integral for $\mathrm{Re}\tilde{\chi}^{zz}(\vb*{0},\omega)$.
The energy cutoff corresponds to the bulk superconducting gap, which is estimated as $\Delta\sim0.1[\mathrm{meV}]$ $(\sim 1[\mathrm{K}])$.
Therefore, our results are applicable in the frequency region below $\hbar\omega\sim0.1[\mathrm{meV}]$ $(\sim 30[\mathrm{GHz}])$.
In addition, we assume that Majorana gap is estimated to be $J_1S\sim0.01[\mathrm{meV}]$ $(\sim 0.1[\mathrm{K}])$.

{\it Discussion.---}Comparing the present results with spin pumping (SP) in a conventional metal-ferromagnet hybrid, the qualitative behaviors are quite different. In conventional metals, spin accumulation occurs due to FMR. In contrast, in the present system, no corresponding spin accumulation occurs due to the Ising anisotropy. Also, in the present calculations, the proximity-induced exchange coupling is assumed to be an isotropic Heisenberg-like coupling. However, in general, the interface interaction can also be anisotropic. Even in such a case, it is no qualitative change in the case of ordinary metals, although a correction term due to anisotropy is added \cite{Matsuo2018}. Therefore, the Ising anisotropy discussed in the present work is a property unique to the Majorana modes and can characterize the Majorana excitations.

Let us comment on the universal nature of the toggling between non-dissipative and dissipative dynamical spin responses observed in our study. Indeed, such toggling becomes universally feasible when the microwave frequency and the energy gap are comparable, and when the Hamiltonian and spin operators are non-commutative, indicating that spin is not a conserved quantity. The non-commutativity can be attributed to the presence of spin-orbit couplings \cite{Thakur2018,Zhou2018,Ominato2018}, and spin-triplet pair correlations \cite{Sigrist1991}.

Microwave irradiation leads to heating within the FI, so that thermally excited magnons due to the heating could influence the DMISR. Phenomena resulting from the heating, which can affect interface spin dynamics, include the spin Seebeck effect (SSE) \cite{Uchida2008}, where a spin current is generated at the interface due to a temperature difference. In hybrid systems of normal metal and FI, methods to separate the inverse spin Hall voltage due to SP from other signals caused by heating have been well studied \cite{iguchi2017}. Especially, it has been theoretically proposed that SP and SSE signals can be separated using a spin current noise measurement \cite{Matsuo2018}. Moreover, SP coherently excites specific modes, which qualitatively differs from SSE induced by thermally excited magnons \cite{Kato2019}. Therefore, even if heating occurs in the FI in our setup, the properties of Majorana Ising spins are expected to be captured. Details of the heating effect on the DMISR will be examined in the near future.

We also mention the experimental feasibility of our theoretical proposals. As we have already explained, the FMR modulation is a very sensitive spin probe. Indeed, the FMR modulation by surface states of 3D topological insulators \cite{Tang2018} and graphene \cite{Patra2012,Tang2013,mendes2015,indolese2018,mendes2019} has been reported experimentally. Therefore, we expect that the enhanced Gilbert damping due to Majorana Ising spin can be observable in our setup when the thickness of the ferromagnetic insulator is sufficiently thin.

Finally, it is pertinent to mention the potential candidate materials where surface Majorana Ising spins could be detectable. Notably, $\mathrm{UTe}_2$ \cite{Ran2019}, $\mathrm{Cu}_x\mathrm{Bi}_2\mathrm{Se}_3$ \cite{Matano2016,Yonezawa2017},  Sr$_x$Bi$_2$Se$_3$ and Nb$_x$Bi$_2$Se$_3$ \cite{Sharma2022-ea} are reported to be in a $p$-wave superconducting state and theoretically can host surface Majorana Ising spins. Recent NMR measurements indicate that $\mathrm{UTe}_2$ could be a bulk $p$-wave superconductor in the Balian-Werthamer state \cite{Matsumura2023}, which hosts the surface Majorana Ising spins with the perpendicular Ising anisotropy, as considered in this work.
$A_x\mathrm{Bi}_2\mathrm{Se}_3$ ($A=$ Cu, Sr, Nb) is considered to possess in-plane Ising anisotropy \cite{Xiong2017}, differing from the perpendicular Ising anisotropy explored in this work.
Therefore, we expect that it exhibits anisotropy different from that demonstrated in this work.

{\it Conclusion.---}We present herein a study of the spin dynamics in a topological superconductor (TSC)-magnet hybrid. Ferromagnetic resonance under microwave irradiation leads to the dynamically induced Majorana Ising spin density on the TSC surface. One can toggle between dissipative and non-dissipative Majorana Ising spin dynamics by adjusting the external magnetic field angle and the microwave frequency.
Therefore, our setup provides a platform to detect and control Majorana excitations.
We expect that our results provide insights toward the development of future quantum computing and spintronics devices using Majorana excitations.

{\it Acknowledgments.---} The authors are grateful to R. Shindou for valuable discussions.
This work is partially supported by the Priority Program of Chinese Academy of Sciences, Grant No. XDB28000000.
We acknowledge JSPS KAKENHI for Grants (Nos. JP20K03835, JP21H01800, JP21H04565, and JP23H01839).

\section*{Supplemental Material}

\subsection{Surface Majorana modes}

In this section, we describe the procedure for deriving the effective Hamiltonian of the surface Majorana modes. We start with the bulk Hamiltonian of a three-dimensional topological superconductor. Based on the bulk Hamiltonian, we solve the BdG equation to demonstrate the existence of a surface-localized solution. Using this solution, we expand the field operator and show that it satisfies the Majorana condition when the bulk excitations are neglected. As a result, on energy scales much smaller than the bulk superconducting gap, the low-energy excitations are described by surface-localized Majorana modes. The above procedure is explained in more detail in the following.
Note that we use $\vb*{r}$ for three-dimensional coordinates and $\rp$ for two-dimensional ones in the Supplemental Material.

We start with the mean-field Hamiltonian given by
\begin{align}
    H_{\mathrm{SC}}=\frac{1}{2}\int d\vb*{r}\hPBdG^\dagger(\vb*{r})\HBdG\hPBdG(\vb*{r}),
    \label{eq_HSC}
\end{align}
with $\vb*{r}=(x,y,z)$.
We consider the Balian-Werthamer (BW) state, in which the pair potential is given by $\Delta_{\hat{\vb*{k}}}=\frac{\Delta}{\kF}\qty(\hat{\vb*{k}}\cdot\vb*{\sigma})i\sigma^y$ with the bulk superconducting gap $\Delta$.
Here, we do not discuss the microscopic origin of the pair correlation leading to the BW state.
As a result, the BdG Hamiltonian $\HBdG$ is given by
\begin{align}
    \HBdG=
    \mqty(
    \ehk-\EF & 0 & -\frac{\Delta}{\kF}\hk_- & \frac{\Delta}{\kF}\hk_x  \\
    0 & \ehk-\EF & \frac{\Delta}{\kF}\hk_x & \frac{\Delta}{\kF}\hk_+   \\
    -\frac{\Delta}{\kF}\hk_+ & \frac{\Delta}{\kF}\hk_x & -\ehk+\EF & 0 \\
    \frac{\Delta}{\kF}\hk_x & \frac{\Delta}{\kF}\hk_- & 0 & -\ehk+\EF  ),
\end{align}
with $\hk_\pm=\hk_y\pm i\hk_z$, $\hat{\vb*{k}}=-i\nabla$, and $\ehk=\frac{\hbar^2\hat{\vb*{k}}^2}{2m}$.
The four component Nambu spinor $\hPBdG(\vb*{r})$ is given by
\begin{align}
    \hPBdG(\vb*{r}):=
    \mqty(
        \hP_{\to}(\vb*{r}) \\
        \hP_{\gets}(\vb*{r}) \\
        \hP_{\to}^\dagger(\vb*{r}) \\
        \hP_{\gets}^\dagger(\vb*{r})
        ),
\end{align}
with the spin quantization axis along the $x$ axis.
The matrices of the spin operators are represented as
\begin{align}
    &\sigma^x=\mqty(1 & 0 \\ 0 & -1), \\
    &\sigma^y=\mqty(0 & 1 \\ 1 & 0), \\
    &\sigma^z=\mqty(0 & -i \\ i & 0).
\end{align}
The fermion field operators satisfy the anticommutation relations
\begin{align}
    &\{\hP_{\sigma}(\vb*{r}),\hP_{\sigma^\prime}(\vb*{r}^\prime)\}=0, \\
    &\{\hP_{\sigma}(\vb*{r}),\hP_{\sigma^\prime}^\dagger(\vb*{r}^\prime)\}
    =\delta_{\sigma\sigma^\prime}\delta(\vb*{r}-\vb*{r}^\prime),
\end{align}
with the spin indices $\sigma,\sigma^\prime=\to,\gets$.

To diagonalize the BdG Hamiltonian, we solve the BdG equation given by
\begin{align}
    \HBdG\vb*{\Phi}(\vb*{r})=E\vb*{\Phi}(\vb*{r}).
    \label{eq_bdg}
\end{align}
We assume that a solution is written as
\begin{align}
    \vb*{\Phi}(\vb*{r})=
    e^{i\kp\cdot\rp}f(z)
    \uvec,
    \label{eq_ss}
\end{align}
with $\vb*{k}_\parallel=(k_x,k_y)$ and $\vb*{r}_\parallel=(x,y)$.
If we set the four components vector to satisfy the following equation (Majorana condition)
\begin{align}
    \mqty( 0 & 0 & 1 & 0 \\ 0 & 0 & 0 & 1 \\ 1 & 0 & 0 & 0 \\ 0 & 1 & 0 & 0)
    \uvec=\pm\uvec,
    \label{eq_PHS}
\end{align}
we can obtain a surface-localized solution.
If we take a positive (negative) sign, we obtain a solution localized on the top surface (bottom surface). As we will consider solutions localized on the bottom surface below, we take a negative sign.
Finally, we obtain the normalized eigenvectors of the BdG equation given by
\begin{align}
    \vb*{\Phi}_{\lambda,\kp}(\vb*{r})=
    \frac{e^{i\kp\cdot\rp}}{\sqrt{L^2}}f_{\kp}(z)
    \vb*{u}_{\lambda,\kp},
\end{align}
with
\begin{align}
    &f_{\kp}(z)=N_{\kp}\sin(k_\perp z)e^{-\kappa z}, \\
    &N_{\kp}=\sqrt{\frac{4\kappa(k_\perp^2+\kappa^2)}{k_\perp^2}}, \\
    &\kappa=\frac{m\Delta}{\hbar^2 \kF}, \\
    &k_{\perp}=\sqrt{\kF^2-{k_{\parallel}^2-\kappa^2}},
\end{align}
and
\begin{align}
    &\vb*{u}_{+,\kp}=
    \mqty(
        u_{+,\to\kp} \\
        u_{+,\gets\kp} \\
        v_{+,\to\kp} \\
        v_{+,\gets\kp}
        )=
    \frac{1}{\sqrt{2}}
    \mqty(
        \sin\frac{\pkp+\pi/2}{2} \\
        -\cos\frac{\pkp+\pi/2}{2} \\
        -\sin\frac{\pkp+\pi/2}{2} \\
        \cos\frac{\pkp+\pi/2}{2} 
        ), \\
    &\vb*{u}_{-,\kp}=
    \mqty(
        u_{-,\to\kp} \\
        u_{-,\gets\kp} \\
        v_{-,\to\kp} \\
        v_{-,\gets\kp}
        )=
    \frac{1}{\sqrt{2}}
    \mqty(
        -\cos\frac{\pkp+\pi/2}{2} \\
        -\sin\frac{\pkp+\pi/2}{2} \\
        \cos\frac{\pkp+\pi/2}{2} \\
        \sin\frac{\pkp+\pi/2}{2} 
        ).
\end{align}
The eigenenergy is given by $E_{\lambda,\kp}=\lambda\Delta k_{\parallel}/\kF$.
We can show that the eigenvectors satisfy
\begin{align}
    \vb*{u}_{-,-\kp}=\vb*{u}_{+,\kp}.
\end{align}
Consequently, the field operator is expanded as
\begin{align}
    \hPBdG(\vb*{r})
    &=\sum_{\kp}
    \qty(
        \gamma_{\kp}
        \frac{e^{i\kp\cdot\rp}}{\sqrt{L^2}}
        +
        \gamma_{\kp}^{\dagger}
        \frac{e^{-i\kp\cdot\rp}}{\sqrt{L^2}}
        ) \notag \\
    &\hspace{1cm}\times f_{\kp}(z)\vb*{u}_{+,\kp}
    +(\mathrm{bulk}{~}\mathrm{modes}),
    \label{eq_mode_expansion}
\end{align}
where $\gamma_{\kp}$ ($\gamma_{\kp}^\dagger$) is the quasiparticle creation (annihilation) operator with the eigenenergy $E_{+,\kp}$.
Substituting the above expression into Eq.\ (\ref{eq_HSC}) with omission of bulk modes and performing the integration in the $z$-direction, we obtain the effective Hamiltonian for the surface states
\begin{align}
    H_{\mathrm{M}}
    =\frac{1}{2}\int d\rp
    \vp^\mathrm{T}(\rp)
    \qty(\hbar v\hat{k}_y\sigma^x-\hbar v\hat{k}_x\sigma^y)
    \vp(\rp),
\end{align}
where $v=\Delta/\hbar\kF$ and we introduced the two component Majorana field operator 
\begin{align}
    \vp(\rp)
    =\mqty(\hp_{\to}(\rp) \\ \hp_{\gets}(\rp)),
\end{align}
satisfying the Majorana condition
\begin{align}
    \hp_{\sigma}(\rp)=\hp_{\sigma}^\dagger(\rp),
\end{align}
and the anticommutation relation
\begin{align}
    \qty{\hp_{\sigma}(\rp),\hp_{\sigma^\prime}(\rp^\prime)}
    =\delta_{\sigma\sigma^\prime}\delta(\rp-\rp^\prime).
\end{align}

The spin density operator of the Majorana mode is given by
\begin{align}
    \vb*{s}(\rp)=
    \vp^\dagger(\rp)
    \frac{\vb*{\sigma}}{2}
    \vp(\rp).
\end{align}
The $x$ component is given by
\begin{align}
    s^x(\rp)
    &=\qty(\hp_{\to}^\dagger(\rp), \hp_{\gets}^\dagger(\rp))
    \mqty(1/2 & 0 \\ 0 & -1/2)
    \mqty(\hp_{\to}(\rp) \\ \hp_{\gets}(\rp)) \notag \\
    &=\frac{1}{2}\qty[\hp_{\to}^\dagger(\rp)\hp_{\to}(\rp)
    -\hp_{\gets}^\dagger(\rp)\hp_{\gets}(\rp)] \notag \\
    &=\frac{1}{2}\qty[\hp_{\to}^2(\rp)-\hp_{\gets}^2(\rp)] \notag \\
    &=0.
\end{align}
In a similar manner, the $y$ and $z$ components are given by
\begin{align}
    s^y(\rp)
    &=\qty(\hp_{\to}^\dagger(\rp), \hp_{\gets}^\dagger(\rp))
    \mqty(0 & 1/2 \\ 1/2 & 0)
    \mqty(\hp_{\to}(\rp) \\ \hp_{\gets}(\rp)) \notag \\
    &=\frac{1}{2}\qty[\hp_{\to}^\dagger(\rp)\hp_{\gets}(\rp)
    +\hp_{\gets}^\dagger(\rp)\hp_{\to}(\rp)] \notag \\
    &=\frac{1}{2}\qty{\hp_{\to}(\rp),\hp_{\gets}(\rp)} \notag \\
    &=0,
\end{align}
and
\begin{align}
    s^z(\rp)
    &=\qty(\hp_{\to}^\dagger(\rp), \hp_{\gets}^\dagger(\rp))
    \mqty(0 & -i/2 \\ i/2 & 0)
    \mqty(\hp_{\to}(\rp) \\ \hp_{\gets}(\rp)) \notag \\
    &=-\frac{i}{2}\qty(\hp_{\to}^\dagger(\rp)\hp_{\gets}(\rp)
    -\hp_{\gets}^\dagger(\rp)\hp_{\to}(\rp)) \notag \\
    &=-i\hp_{\to}(\rp)\hp_{\gets}(\rp),
\end{align}
respectively.
As a result, the spin density operator is given by
\begin{align}
    \vb*{s}(\rp)=\qty(0,0,-i\hp_{\to}(\rp)\hp_{\gets}(\rp)).
\end{align}
One can see that the spin density of the Majorana mode is Ising like.

\subsection{Majorana Ising spin dynamics}

In this section, we calculate the Ising spin density induced on the TSC surface by the proximity coupling $\Hex$.
$\Hex$ consists of two terms, $\HZ$ and $\HT$. $\HZ$ leads to the static spin density and $\HT$ leads to the dynamic spin density. First, we calculate the static spin density. Next, we calculate the dynamic spin density.

The total spin density operator is given by
\begin{align}
    s^z_{\mathrm{tot}}
    &=\int d\rp s^z(\rp).
\end{align}
The statistical average of the static spin density is calculated as
\begin{align}
    \expval{s^z_{\mathrm{tot}}}
    &=-\sum_{\kp}
    \frac{M}{2E_{\kp}}\qty[f(E_{\kp})-f(-E_{\kp})] \notag \\
    &\to -\qty(\frac{L}{2\pi\hbar v})^2
    \int_{M}^{\Delta} EdE
    \int_0^{2\pi}d\phi\frac{M}{2E}
    \qty[f(E)-f(-E)] \notag \\
    &=-\int_{0}^{\Delta} dE
    D(E)
    \frac{f(E)-f(-E)}{2E}M.
\end{align}
At the zero temperature limit $T\to0$, the static spin density is given by
\begin{align}
    \expval{s^z_{\mathrm{tot}}}
    &=\frac{1}{2}
    \frac{L^2}{2\pi(\hbar v)^2}\qty(\Delta-M)M
    \approx\tilde{\chi}^{zz}_0(\vb*{0},0)M,
\end{align}
where $\tilde{\chi}^{zz}_0(\vb*{0},0)=D(\Delta)/2$ and we used $\Delta\gg M$.

The dynamic spin density is given by the perturbative force
\begin{align}
    \HT(t)=\int d\rp s^z(\rp)F(\rp,t),
\end{align}
where $F(\rp,t)$ is given by
\begin{align}
    F(\rp,t)
    &=-\sin\vartheta\sum_n J(\rp,\vr_n) \expval{S_n^X(t)} \notag \\
    &\approx-\sin\vartheta J_1S\frac{\gamma h_{\mathrm{ac}}}{\sqrt{(\omega-\gamma h_{\mathrm{dc}})^2+\alpha^2\omega^2}}\cos\omega t \notag \\
    &=:F\cos\omega t.
\end{align}
The time dependent statistical average of the Ising spin density, to the first order of $J_1S$, is given by
\begin{align}
    &\int d\rp\expval{s^z(\rp,t)} \notag \\
    &=
    \int d\rp
    \int d\rp^\prime
    \int dt^\prime
    \chi^{zz}(\rp-\rp^\prime,t^\prime)
    F(\rp^\prime,t-t^\prime) \notag \\
    &=\mathrm{Re}\qty[\tilde{\chi}^{zz}(\vb*{0},\omega)Fe^{-i\omega t}] \notag \\
    &\approx\mathrm{Re}\qty[\tilde{\chi}^{zz}_{0}(\vb*{0},\omega)]F\cos\omega t,
\end{align}
where we used $\mathrm{Re}\tilde{\chi}^{zz}_{0}(\vb*{0},\omega)\gg\mathrm{Im}\tilde{\chi}^{zz}_{0}(\vb*{0},\omega)$.
The real part of $\tilde{\chi}^{zz}(\vb*{0},\omega)$ is given by
\begin{align}
    \mathrm{Re}\tilde{\chi}^{zz}(\vb*{0},\omega)
    &=-\mathcal{P}\int dED(E)\frac{E^2-M^2}{2E^2}
    \frac{f(E)-f(-E)}{2E+\hbar\omega},
\end{align}
where $\mathcal{P}$ means the principal value.
When the integrand is expanded with respect to $\omega$, the lowest order correction term becomes quadratic in $\omega$. In the frequency range considered in this work, this correction term is significantly smaller compared to the static spin susceptibility $\mathrm{Re}\tilde{\chi}^{zz}(\vb*{0},0)$. Therefore, the spin susceptibility exhibits almost no frequency dependence and remains constant as a function of $\omega$.
The imaginary part of $\tilde{\chi}^{zz}(\vb*{0},\omega)$ is given by
\begin{align}
    &\mathrm{Im}\tilde{\chi}^{zz}(\vb*{0},\omega) \notag \\
    &=\pi D(\hbar\omega/2)
    \frac{(\hbar\omega/2)^2-M^2}{2(\hbar\omega/2)^2}
    \qty[
        f(-\hbar\omega/2)-f(\hbar\omega/2)
    ].
\end{align}

\subsection{FMR modulation due to the proximity exchange coupling}

In this section, we provide a brief explanation for the derivation of the FMR modulations $\delta h$ and $\delta\alpha$. The FMR modulations can be determined from the retarded component of the magnon Green's function, which is given by
\begin{align}
&\tilde{G}^R(\vk,\omega)
=\frac
{2S/\hbar}
{\omega-\omega_{\vk}+i\alpha\omega
-(2S/\hbar)\Sigma^R(\vk,\omega)
},
\end{align}
where we introduce the Gilbert damping constant $\alpha$ phenomenologically.
In the second-order perturbation calculation with respect to $\HT$, the self-energy is given by
\begin{align}
    \Sigma^R(\vk,\omega)=-\qty(\frac{\sin\vartheta}{2})^2
    \sum_{\qp}|\tilde{J}(\qp,\vk)|^2\tilde{\chi}^{zz}(\qp,\omega),
    \label{eq_self_energy_J2chi}
\end{align}
where $\tilde{J}(\qp,\vb*{0})$ is given by
\begin{align}
    \tilde{J}(\qp,\vk)=\frac{1}{L^2\sqrt{N}}\int d\rp\sum_nJ(\rp,\vb*{r}_n)e^{i(\qp\cdot\rp+\vk\cdot\vb*{r}_n)}
\end{align}

The pole of $\tilde{G}^R(\vk,\omega)$ signifies the FMR modulations, including both the frequency shift and the enhanced Gilbert damping. These are given by
\begin{align}
    \delta h=
    \frac{2S}{\gamma\hbar}
    \mathrm{Re}\Sigma^R(\vb*{0},\omega),
    \hspace{2mm}
    \delta\alpha=
    -\frac{2S}{\hbar\omega}
    \mathrm{Im}\Sigma^R(\vb*{0},\omega).
    \label{eq_delta_H_delta_alpha}
\end{align}
From the above equations and Eq.~(\ref{eq_self_energy_J2chi}), it is apparent that FMR modulations provide information regarding both the properties of the interface coupling and the dynamic spin susceptibility of the Majorana modes.

The form of matrix element $\tilde{J}(\qp,\vb*{0})$ depends on the details of the interface. In this work, we assume the specular interface. $|\tilde{J}(\qp,\vb*{0})|^2$ is given by
\begin{align}
    |\tilde{J}(\qp,\vb*{0})|^2
    =
    \frac{J_1^2}{N}\delta_{\qp,\vb*{0}}.
    \label{eq_matrix_element_model}
\end{align}
Using Eq.\ (\ref{eq_matrix_element_model}), the self-energy for the uniform magnon mode is given by
\begin{align}
    &\Sigma^R(\vb*{0},\omega)
    =
    -\qty(\frac{\sin\vartheta}{2})^2
    \frac{J_1^2}{N}
    \tilde{\chi}^{zz}(\vb*{0},\omega).
\end{align}

\bibliography{ref}

\end{document}